\documentclass[prl,showpacs,aps,twocolumn,amsmath,amssymb,amsfonts]{revtex4}
\usepackage[english]{babel}
\newcommand{\pd}{\partial}

\newcommand{\beq}{\begin{equation}}
\newcommand{\eeq}{\end{equation}}
\newcommand{\bea}{\begin{eqnarray}}
\newcommand{\eea}{\end{eqnarray}}

\begin{document}
\selectlanguage{english}

\title{Black ring with two angular momenta}

\author{\firstname{A.~A.}~\surname{Pomeransky}}
\email{a.a.pomeransky@inp.nsk.su}
\author{\firstname{R.~A.}~\surname{Sen'kov}}
\email{r.a.senkov@inp.nsk.su}

\affiliation{Budker Institute of Nuclear Physics, 630090,
Novosibirsk, Russia,\\
and Novosibirsk University}


\begin{abstract}
General regular black ring solution with two angular momenta is
presented, found by the inverse scattering problem method. The mass,
angular momenta and the event horizon volume are given explicitly as
functions of the metric parameters.
\end{abstract}
 \pacs{04.20.Jb, 04.50.+h, 04.70.Bw, 05.45.Yv}

\maketitle

The study of gravitation in space-times of various dimensions is
required by string theory. Black holes are the objects of crucial
importance for gravitation. One of their characteristic features is
the small number of parameters they depend on. In four dimensions,
according to the no-hair theorems, the only allowed parameters are
mass, angular momentum and gauge charges. The uncharged solution was
found by Kerr \cite{Kerr:1963}, it has an event horizon of spherical
topology. Its higher-dimensional analogue \cite{Myers:1986} was
known already for some time, when the five-dimensional black ring
solution was discovered \cite{Emparan:2001}, having the event
horizon topology $S^1\times S^2$. Then it became clear that the
family of black holes has a richer structure in more space-time
dimensions. Subsequently, a lot of attention was focused on black
rings in 5D supergravity (see \cite{Emparan:2006} for a review).

A body in the four-dimensional space can rotate in two mutually
orthogonal planes and thus has two independent angular momentum
parameters. The black ring  solution presented in
\cite{Emparan:2001} has only one angular momentum, the other being
equal to zero, and obviously can not be the most general one. In
this article we present the general regular black ring solution of
five-dimensional general relativity with two independent angular
momenta. The solution is constructed using the complete
integrability of this system by the inverse scattering problem
method. We use Belinski-Zakharov method
\cite{Belinski:1978,Belinski:1979} of adding solitons to a seed
solution. The effectiveness of this approach for finding
higher-dimensional black hole solutions was tested in
\cite{Pomeransky:2005} by rederiving the 5D Myers-Perry solution. To
apply this technique to black rings it was necessary to know an
appropriate seed solution. We find it by combining the results of
\cite{Mishima:2005,Mishima:2006,Tomizawa:2005,Tomizawa:2006,Figueras:2005}.
We also used extensively \cite{Harmark:2004} at all stages of our
work.

 The family of  2-soliton solutions, that we have found, contains not
 only regular black rings, but also black rings with the conical
 singularity and the Dirac string singularity of arbitrary
 magnitudes. We do not give here the expressions for these singular
 metrics, because they are somewhat longer than in the regular case.
 They will be presented in a forthcoming paper together with a more
 detailed derivation of the solutions.
 The main result of this paper is the following metric of the general regular black ring
 (we use a mostly minus signature):
 \bea\label{metrica}
 &ds^2&=\frac{H(y,x)}{H(x,y)}(dt+\Omega)^2+\frac{F(x,y)}{H(y,x)}d\phi^2+2\frac{J(x,y)}{H(y,x)} d\phi d\psi\nonumber\\
  &-&\frac{F(y,x)}{H(y,x)}d\psi^2-\frac{2 k^2 H(x,y)}{(x-y)^2(1-\nu)^2}(\frac{dx^2}{G(x)}-\frac{dy^2}{G(y)}).
 \eea
 The metric above depends only on two of the coordinates:
 $-1\leq x \leq 1$ and $-\infty < y <-1$.
 They are close analogues of the coordinates used in \cite{Figueras:2005}.
  The metric is independent of time $-\infty<t<+\infty$, and angles $0<\phi,\psi<2\pi$.
 The one-form $\Omega$ is given by
 \bea\label{omega}
 &\Omega&=-\frac{2 k \lambda \sqrt{(1+\nu )^2-\lambda ^2}}{H(y,x)}( (1-x^2) y \sqrt{\nu}d\psi+
 \frac{(1+y)}{(1-\lambda +\nu)}\nonumber\\
 &\times&(1+\lambda -\nu +x^2 y \nu (1-\lambda-\nu)+2\nu x(1-y))\,d\phi),
 \eea
 and the functions $G$, $H$, $J$, $F$ are defined as:
\begin{widetext}
 \bea\label{functions}
 G(x)&=&(1-x^2)(1+\lambda x+\nu x^2)\,,\nonumber\\
 H(x,y)&=& 1+\lambda ^2-\nu^2+2\lambda\nu (1-x^2)y
      +2x\lambda(1-y^2\nu^2)+ x^2 y^2 \nu(1-\lambda^2-\nu^2)\,,\nonumber\\
 J(x,y)&=&\frac{2 k^2 (1-x^2) (1-y^2) \lambda  \sqrt{\nu}}{(x-y) (1-\nu)^2}
 \,(1+\lambda ^2 -\nu ^2
  + 2 (x+y) \lambda  \nu-x y \nu (1-\lambda ^2-\nu^2)),\\
 F(x,y) &=& \frac{2 k^2}{(x-y)^2 (1-\nu)^2} (G(x) (1-y^2)\left(\left((1-\nu)^2-\lambda ^2\right)
  (1+\nu )+y \lambda (1-\lambda ^2+2 \nu -3 \nu ^2)\right)+ G(y) (2 \lambda ^2\nonumber\\
    &+& x \lambda ((1-\nu )^2+\lambda ^2)
   + x^2\left((1-\nu )^2-\lambda ^2\right) (1+\nu)+x^3\lambda(1-\lambda^2-3\nu^2+2\nu^3)
   - x^4 (1-\nu ) \nu (-1+\lambda ^2+\nu ^2))).\nonumber
   \eea
 \end{widetext}
 For a regular black ring, parameters $\lambda$  and $\nu$ have to
 satisfy the constraints $0\leq\nu<1$ and
 $2\sqrt{\nu}\leq\lambda<1+\nu$. These constraints can be derived
 from the existence of the horizons (both roots of the equation
 \beq\label{horizon}
 1+\lambda y+\nu y^2=0
 \eeq
 have to be $<-1$), reality of the
 metric ($\nu \geq 0$) and positivity of the black ring mass
 ($\lambda>0$, see below). One recovers the Emparan-Reall ring
 rotating only in the $\phi$ direction by setting $\nu=0$. The black
 ring has a regular event horizon situated at $y=y_h$, where
 \beq
 y_h=\frac{-\lambda+\sqrt{\lambda^2-4\nu}}{2\nu}
 \eeq
  is the root of the Eq. (\ref{horizon}) with the smallest absolute value.
 The event horizon volume is equal to
 \beq\label{volume}
  V_h=-\frac{32\pi^2
  k^3(1+\lambda+\nu)\lambda}{(y_h-1/y_h)(1-\nu)^2}.
 \eeq
 The extreme rotating black ring is obtained in the limit
 $\lambda=2\sqrt{\nu}$, when both roots of the Eq. (\ref{horizon}) coincide.

 Apart from $(x,y)$ coordinates, the so called canonical coordinates $\rho$ and $z$ constitute
 another important choice \cite{Belinski:1978,Belinski:1979,Harmark:2004}.
 They are introduced in the following way. The metric is decomposed in two square blocks: one is the $3\times
 3$ matrix $g_{ab}$, $a,b=t,\phi,\psi$ and the other is the $2\times
 2$ matrix corresponding to $x,y$ coordinates. The function
 $\rho=\sqrt{\det{g_{ab}}}$ is taken as one of the new coordinates,
 while the choice of the other coordinate $z$ is restricted by the
 condition that the $2\times 2$ part of the metric is conformally
 flat in $(\rho,z)$ coordinates. One has
 \bea
 \rho^2&=& -\,\frac{4k^4 G(x)G(y)}{(x-y)^4 (1-\nu)^2}\,,\\
 z&=&\frac{k^2 (1-x y) (2+(x+y)\lambda+2x y\nu)}{(x-y)^2 (1-\nu)}\,.\nonumber
 \eea

 It can be checked directly that the metric (\ref{metrica}) satisfies
 sourceless Einstein equations (that is its Ricci tensor $R_{\mu\nu}$ vanishes).
 The Einstein equations in our case reduce \cite{Harmark:2004}  to
 a $3\times 3$ matrix equation:
 \beq
 \pd_i (\rho g^{ab}\pd_i g_{bc})=0, \;\;\; i=\rho\,,\;z
 \eeq
 and a pair of equations for the conformal factor $f$:
 \bea\label{factor}
 \pd_\rho \ln f &=&
 -\rho^{-1}+\frac{\rho}{4}(g_{ab,\rho}g_{cd,\rho}-g_{ab,z}g_{cd,z})g^{ac}g^{bd}\,,\nonumber\\
 \pd_z \ln f &=& \frac{\rho}{2}g_{ab,\rho}g_{cd,z}g^{ac}g^{bd}\,,\;\;\;f=-g_{\rho\rho}=-g_{zz}.
 \eea

 The black ring space-time (\ref{metrica}) is asymptotically flat, with the spacial infinity
 located at the point $(-1,-1)$ in the $(x,y)$ plane. The
 asymptotic flatness becomes clear when the new coordinates
 $(r_1,r_2)$  are introduced by the set of equations: $\rho=r_1 r_2$,
 $z=(r_1^2-r_2^2)/2$, together with their combination $r$: $r^2=r_1^2+r_2^2=2\sqrt{\rho^2+z^2}$. In the new
 coordinates the flat space metric has the form:
 \beq
 dl^2=dr_1^2+dr_2^2+r_1^2 d\phi^2+r_2^2 d\psi^2.
 \eeq
 The asymptotics of the metric components at
 spacial infinity has the form \cite{Harmark:2004}:
  \beq
   g_{tt}=1-\frac{8 G_N M}{3\pi r^2},\; g_{t\phi}=\frac{4 G_N S_\phi r_1^2}{\pi r^4},\;
   g_{t\psi}=\frac{4 G_N S_\psi r_2^2}{\pi r^4}.
  \eeq
  Comparing this with the actual asymtotics of the metric (\ref{metrica})--(\ref{functions})
  we find the black ring's mass and angular momenta:
 \bea
 M&=&\frac{3 k^2 \pi \lambda }{G_N (1-\lambda +\nu )},\;\;\;\;
 S_\psi=\frac{4 k^3 \pi  \lambda  \sqrt{\nu } \sqrt{(1+\nu )^2-\lambda^2}}{G_N (1-\nu )^2 (1-\lambda +\nu)},\nonumber\\
  S_\phi&=&\frac{2 k^3 \pi  \lambda  \left(1+\lambda -6 \nu +\lambda  \nu +\nu ^2\right)
 \sqrt{(1+\nu )^2-\lambda ^2}}{G_N (1-\nu )^2 (1-\lambda +\nu )^2},
 \eea
 where $G_N$ is the Newton constant.

 We derived the general black ring metric by adding two solitons to
 a specially chosen  background. The  way to find the seed metric  was
 prompted to us by works \cite{Mishima:2005,Tomizawa:2005},
 where the black ring with conical singularity,
 rotating along $\psi$ angle was found for the first time
 by adding two solitons to a static seed metric. This solution is
 complementary to the Emparan-Reall black ring, which rotates in the other plane
 (along $\phi$ angle). The seed solution of \cite{Mishima:2005,Tomizawa:2005}
 can be obtained by removing two solitons from the static black ring.
 The operation of removing solitons is described in
 \cite{Pomeransky:2005}, and in the present case it consists in
 dividing $g_{\psi\psi}$ component of the metric by a certain simple
 function. Then, it is natural to suppose, that by removing two solitons from
 the Emparan-Reall black ring, one can obtain the seed metric we are looking
 for. Really, adding to it back a pair of solitons (with more general soliton parameters,
 of course) will make it rotate in the other plane too. We have
 found that this is indeed the case. It is worth to note that
 the regular black ring with two angular momenta comes from
 the seed solution, which is obtained from the regular Emparan-Reall ring
 (without conical singularity). Another fortunate fact is  that to obtain the regular solution,
 one should set equal to zero one of the constant parameters for each soliton,
 which makes the calculation easier.

 There remained however the following significant technical difficulty at this
 stage. In the inverse scattering method one needs to know the solution of a linear system
 of differential equations, the $3 \times 3$ matrix $\psi_0(\Lambda)$,
 which corresponds to the seed metric (\cite{Belinski:1978,Belinski:1979}, see also \cite{Pomeransky:2005}).
 The black ring metric is not diagonal even if the ring rotates in a single plane
 only, and thus our seed is not diagonal too. It is hard to find the solution of the linear system
 $\psi_0(\Lambda)$ when the corresponding metric is not a diagonal matrix.  Fortunately, this problem
 was solved in \cite{Mishima:2006,Tomizawa:2006}, where the
 Emparan-Reall ring was reproduced by the inverse scattering
 method.  This automatically gives us the corresponding
 $\psi_0(\Lambda)$ matrix. While originally two-soliton solutions were
 used in \cite{Mishima:2006,Tomizawa:2006}, it is sufficient to consider
 one-soliton solutions only, which significantly simplifies the calculations.

 After we have obtained the solution, an important and difficult task was to
 rewrite it in as simple form as possible.
 Initially, the Belinski-Zakharov method gives the metric in the
 canonical coordinates $(\rho,z)$. Then, the metric can be cast
 in a considerably simpler form (\ref{metrica})
 by going over to the coordinates $(x,y)$. We have found this transformation
 by comparing the metric of \cite{Mishima:2005} in coordinates $(\rho,z)$ \cite{Tomizawa:2005}
 with its much simpler form, written in coordinates  $(x,y)$ in \cite{Figueras:2005}.
 The coordinates $(x,y)$ are related to the C-metric coordinates used in \cite{Harmark:2004}
 by a M\"obius linear fractional transformation with fixed points $\pm 1$. The single remaining
 parameter of this transformation is then chosen in such a way, that the
 $g_{\psi\psi}$ component of the metric has the simple form (\ref{omega}).

 After this work was completed e-print \cite{Kudoh:2006} has
 appeared, where black rings with two angular momenta are studied
 by perturbative and numerical methods. It would be very interesting
 to compare the results of \cite{Kudoh:2006} with the corresponding limits
 of our exact solution.

 \begin{acknowledgments}
 We are grateful to Anatoly Pomeransky and Roman Lee for many useful
 discussions. The investigation was supported in part by the Russian
 Foundation for Basic Research through Grant No. 05-02-16627.
 \end{acknowledgments}


\begin{thebibliography}{99}
 \bibitem{Kerr:1963} R.~P. Kerr, Phys. Rev. Lett. {\bf 11}, 237 (1963).
 \bibitem{Myers:1986} R.~C. Myers and M.~J. Perry, Ann. Phys. {\bf 172}, 304 (1986).
 \bibitem{Emparan:2001} R. Emparan and H.~S. Reall, Phys. Rev. Lett. {\bf 88}, 101101 (2002), hep-th/0110260.
 \bibitem{Emparan:2006} R. Emparan and H.~S. Reall, Class. Quant. Grav. {\bf 23}, R169 (2006), hep-th/0608012.
 \bibitem{Belinski:1978} V. A. Belinski and  V. E. Zakharov, Sov. Phys. JETP {\bf 48}, 985 (1978).
 \bibitem{Belinski:1979} V. A. Belinski and  V. E. Zakharov, Sov. Phys. JETP {\bf 50}, 1 (1979).
 \bibitem{Pomeransky:2005} A. A. Pomeransky, Phys. Rev. {\bf D73}, 044004 (2006), hep-th/0507250.
 \bibitem{Mishima:2005} T. Mishima and H. Iguchi, Phys. Rev. {\bf D73}, 044030 (2006), hep-th/0504018.
 \bibitem{Mishima:2006} T. Mishima and H. Iguchi, Phys. Rev. {\bf D73}, 121501 (2006), hep-th/0604050.
 \bibitem{Tomizawa:2005} S. Tomizawa, Y. Morisawa, and Y. Yasui, Phys. Rev. {\bf D73}, 064009 (2006), hep-th/0512252.
 \bibitem{Tomizawa:2006} S. Tomizawa and M. Nozawa, Phys. Rev. {\bf D73}, 124034 (2006), hep-th/0604067.
 \bibitem{Figueras:2005} P. Figueras, JHEP {\bf 0507}, 039 (2005), hep-th/0505244.
 \bibitem{Harmark:2004} T. Harmark, Phys. Rev. {\bf D70}, 124002 (2004), hep-th/0408141.
 \bibitem{Kudoh:2006} H. Kudoh, gr-qc/0611136.

\end{thebibliography}
\end{document}